\documentclass[preprint]{sig-alternate-05-2015}

\usepackage{array}
\usepackage{graphicx}
\usepackage{url}

\newcolumntype{C}[1]{>{\centering\let\newline\\\arraybackslash\hspace{0pt}}p{#1}}

\def\sharedaffiliation{
\end{tabular}
\begin{tabular}{c}
}

\begin{document}

\title{Towards co-designed optimizations in parallel frameworks: A MapReduce case study}

\numberofauthors{3}

\author{
  \alignauthor 
    Colin Barrett
  \alignauthor 
    Christos Kotselidis
  \alignauthor
    Mikel Luj{\'a}n
  \sharedaffiliation
    \affaddr{The University of Manchester}\\
    \affaddr{Oxford Road, Manchester, M13 9PL, UK}\\
    \email{\mbox{\{colin.barrett, christos.kotselidis, mikel.lujan\}@manchester.ac.uk}}
}

\maketitle

\begin{abstract}

The explosion of Big Data was followed by the proliferation of numerous complex parallel software stacks whose aim is to tackle the challenges of data deluge.
A drawback of a such multi-layered hierarchical deployment is the inability to maintain and delegate vital semantic information between layers in the stack.
Software abstractions increase the semantic distance between an application and its generated code.
However, parallel software frameworks contain inherent semantic information that general purpose compilers are not designed to exploit.

This paper presents a case study demonstrating how the specific semantic information of the MapReduce paradigm can be exploited on multicore architectures.
MR4J has been implemented in Java and evaluated against hand-optimized C and C++ equivalents.
The initial observed results led to the design of a \textit{semantically aware} optimizer that runs automatically without requiring modification to application code.

The optimizer is able to speedup the execution time of MR4J by up to 2.0x.
The introduced optimization not only improves the performance of the generated code, during the map phase, but also reduces the pressure on the garbage collector.
This demonstrates how semantic information can be harnessed without sacrificing sound software engineering practices when using parallel software frameworks.

\end{abstract}

\ccsdesc[500]{Software and its engineering~Frameworks}
\ccsdesc[300]{Software and its engineering~Runtime environments}
\ccsdesc[100]{Software and its engineering~Source code generation}

\printccsdesc

\section{Introduction}

Parallel software frameworks facilitate the separation of concerns between functionality and parallelism.
Cilk \cite{Blumofe:1995}, X10 \cite{Charles:2005} and the Fork/Join pool from the Java Development Kit (JDK) \cite{JavaSE:2015}, among others, provide an interface to explicitly parallelize applications by abstracting away direct interactions with the underlying architectures.
The challenge with explicit parallelism is to create an efficient application; a pareto-optimal point between introduced overheads and achieved performance.
The need for increased productivity has lead to a boom in the number and popularity of parallel frameworks.
These are deployed in various combinations creating rich multi-layered software stacks.
Consequently, the flow of information from the application down to the code generator passes a number of intermediate steps in which the information is constantly abstracted and/or optimized.

Compiler optimization principles such as loop invariant code motion, inlining and scalar replacement \cite{Muchnick:1997}, were developed in order to enhance performance by achieving better machine code quality.
They rely on control and data dependencies contained within an intermediate representation.
Applications based on parallel frameworks are composed from smaller tasks with few explicit data dependencies.
However, many of these frameworks contain the semantic information required to infer these dependencies and perform optimizations that are not detected in the original form.
Dynamic compilers are not designed to exploit the inferred data dependencies inherent in parallel software frameworks.
Furthermore, since they operate on fine-grain abstractions (methods), they do not have a full picture of the application and thus cannot infer valuable knowledge for further optimizations.

This paper provides a case study where both the application and the parallel framework semantics are examined together in order to detect missing optimization opportunities.
After identifying such opportunities, we apply familiar compilation techniques on current parallel frameworks in order to bridge the semantic gap between application logic and the programmability of frameworks in a co-designed manner.
The semantics of the framework were used to design an optimizer that improved the performance of applications running on top of it with no user involvement while maintaining software engineering principles.
This paper makes the following contributions:

1) Introduces MR4J, a lightweight Java based MapReduce framework for multicore architectures.

2) Demonstrates a co-designed optimizer where code transformations are automatically applied to enhance performance of running applications.

3) Provides an in-depth comparative performance analysis of MR4J and other shared-memory MapReduce frameworks.

\section{MapReduce: A Case Study}

MapReduce is a popular framework for regular data parallel applications originally designed for running on distributed systems.
Due to its efficiency and scalability recent efforts have provided implementations for multicore architectures.
As many `cloud' applications do not necessarily require any more resources than a single node \cite{Appuswamy:2013}, multicore implementations offer a practical alternative.
In general, MapReduce requires the provision of two tasks, each with a well-defined objective; making it an ideal framework to demonstrate optimizations based on implicit semantic information of an abstraction.

This paper implements MR4J, a lightweight implementation of MapReduce for multicore architectures to demonstrate the applicability of the abstraction within the Java programming language.
MR4J was designed to supplement existing frameworks in order to take advantage of the rich Java Application Programming Interface (API), the Virtual Machine (JVM) portability across different architectures, and recent efforts in compiler abstractions \cite{Duboscq:2013, Wurthinger:2013}.

Managed runtime languages, in particular Java, employ automatic memory management techniques, in the means of Garbage Collectors (GC) \cite{Jones:2011} that alleviate the users from the difficult and error-prone task of manual memory management.
This fact, however, introduces a performance overhead, a characteristic of managed runtime languages and is tackled by the optimizer introduced in this paper in the context of MR4J.
The performance of MR4J is evaluated against the state-of-the-art C and C++ equivalent frameworks in order to quantify the performance/productivity trade-off.
The design and implementation of the optimization creates new opportunities in defining a standard methodology for applying similar techniques to other parallel frameworks.

\subsection{What is MapReduce?}

MapReduce transforms input data into a collection of (key, value) pairs.
In order to achieve this, the input is split and individually passed as an argument to the map method.
The \textit{map} method emits intermediate (key, value) pairs that are collected by the framework and consequently grouped for the reduce phase.
The \textit{reduce} method combines all the intermediate values associated with each key into the (key, value) pairs returned as the result.
The input data is assumed independent, when split, so the benefit of this approach is that execution of each map and reduce method can be performed in parallel.

\begin{figure}
\centering
\includegraphics[trim = 0mm 50mm 0mm 0mm, clip, width=80mm]{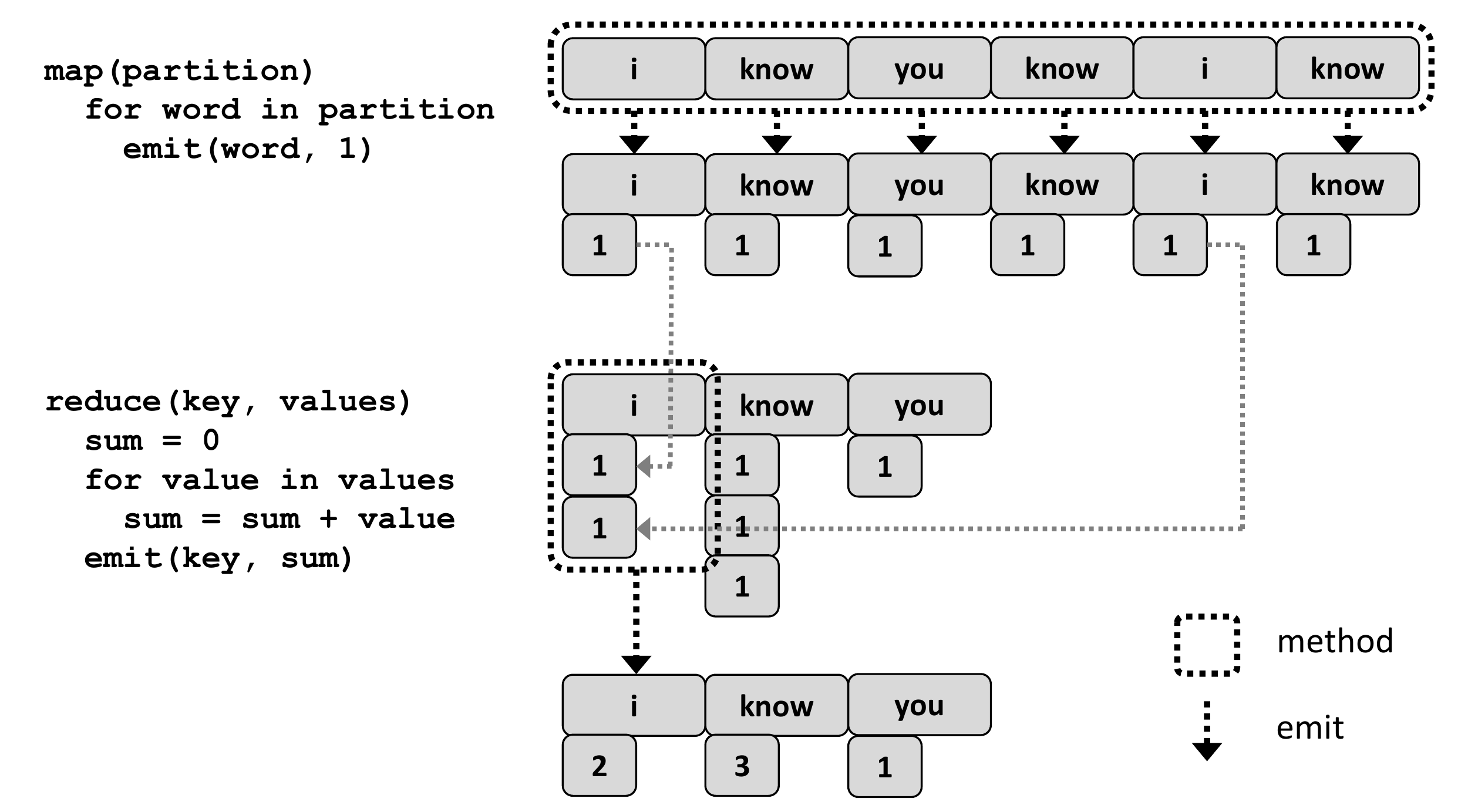}
\caption{Illustration of MapReduce used for a word count application used as a running example.}
\label{fig:mapreduce}
\end{figure}

\begin{figure}
{\scriptsize
\begin{verbatim}
public class WordCount {
  static final Pattern WORD = Pattern.compile("[A-Z][A-Z']*");
  final Mapper<S, S, I> mapper = new Mapper<S, S, I>() {
    public void map(S input, Emitter<S, I> emitter) {
      Matcher words = WORD.matcher(input.toUpperCase());
        while (words.find()) {
          emitter.emit(words.group(), 1);
        }
     }
  };
  final Reducer<S, I> reducer = new Reducer<S, I>() {
    public void reduce(S key,
                       List<I> values,
                       Emitter<S, I> emitter) {
      int sum = 0;
      for (I value : values) {
        sum += value;
      }
      emitter.emit(key, sum);
    }
  };
  public List<KeyValue<S, I>> run(List<S> input) {
    MapReduce<S, S, I> mrj = new MapReduce<>(mapper, reducer);
      return mrj.run(input);
  }
} \end{verbatim} }
\caption{Implementation of the word count application using MR4J.}
\label{fig:implementation}
\end{figure}

Figure \ref{fig:mapreduce} contains a pseudo-code example of the map and reduce methods for a word count application and its information flow respectively.
Figure \ref{fig:implementation} contains a working implementation of this running example based on MR4J, introduced in this paper.
As depicted in Figure \ref{fig:mapreduce}, the map method receives a sentence as an argument and splits it into individual words (each with an initial count of one).
Each word is then emitted into the framework where the individual counts are collected for each unique word.
The word and its counts are consequently passed to the reduce method as its arguments.
The reduce method, in turn, accumulates the values to form the final count for the word which is emitted as the result.

\subsection{Related Frameworks}

The application of MapReduce as an abstraction spans web analytics \cite{Dean:2008}, machine learning \cite{Mahout:2013}, and databases \cite{MongoDB:2013}.
Due to its flexibility in targeting different hardware architectures, there is a wide range of implementations ranging from distributed to multicore deployments.

\subsubsection{Distributed Networks (Clusters/Clouds)}

Google coined the MapReduce name \cite{Dean:2008} and took the first steps to popularize the framework by providing an API to automatically split and distribute the input data and the execution across a cluster of processing nodes.
The API, written in C++, hides many aspects of the underlying parallelism (e.g. the scheduling, data distribution,
and fault tolerance) and relies on the Google File System (GFS) to distribute the data \cite{Ghemawat:2003}.
Dean \textit{et~al.} further refined the abstraction by adding a new method to \textit{combine} intermediate values on a processing node.
This partially reduces values associated with each key and minimizes data transfers before the reduce phase.

Hadoop \cite{Hadoop:2013}, an open source implementation of the MapReduce framework in Java, is implemented in a similar manner to the Google MapReduce framework, including the same refinements.
Hadoop uses Java interfaces to define the map and reduce methods using generic types; allowing flexible, yet typed, parameters for the (key, value) pairs.
It is also possible to configure the framework to utilize multicore systems.

\subsubsection{Multicore Architectures}

Phoenix 1.0 \cite{Ranger:2007}, Phoenix 2.0 \cite{Yoo:2009} (written in C) and Phoenix++ 1.0 \cite{Talbot:2011} (written in C++) utilize the principles of the MapReduce framework and substitute the communication strategies of clusters with shared-memory buffers.
This approach replaces worker nodes with threads to execute the tasks minimizing the overheads of the framework.
The principle aim of the project is to provide ``an efficient implementation on shared-memory systems that demonstrates its feasibility'' \cite{Ranger:2007}.
The use of threads and shared-memory enables optimizations for data locality and, with some risk to correctness, shared mutable state.
The popularity of MapReduce has encouraged implementations for different architectures and because of the complexity of memory management, the API is restrictive and closer to the original concept proposed by Dean \textit{et~al.} \cite{Dean:2008}.

\begin{figure}
\centering
\includegraphics[trim = 0mm 75mm 0mm 0mm, clip, width=80mm]{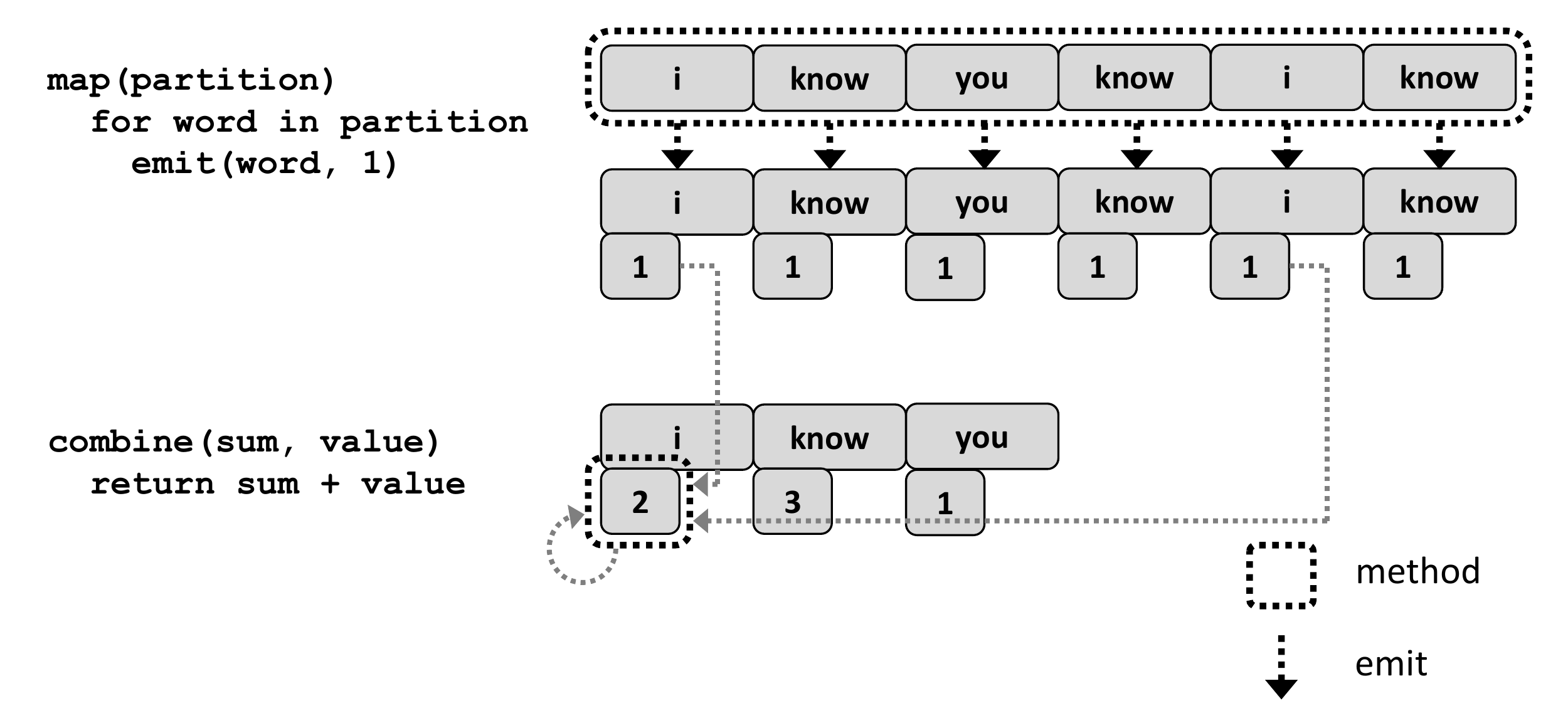}
\caption{Illustration of the creation and management of (key, value) pairs in MapReduce for the word counting example with a combiner replacing the reduce method.}
\label{fig:mapcombine}
\end{figure}

\subsection{Performance vs. Programmability}

A common feature of all the existing MapReduce frameworks is the acknowledgment that, without modifications to its purest form, performance is limited.
For example, the combiner method exists to reduce the size of data in the intermediate (key, value) collection.
As the implementation and further manual tuning remain external to the framework; there is an assumption of familiarity and a knowledge of parallel programming, a discipline which is known to be challenging.

Implementation of combining functions in the Phoenix frameworks improves the performance but introduces a deterioration in programmability.
Phoenix adds a new function prototype that, when implemented and supplied as an argument to the framework, incrementally combines intermediate values in a small buffer to a single value in order to prevent the allocation of new memory for the collector.
Although it improves execution time, it often duplicates the code written by the user.
This issue is further compounded by the use of void pointers for `generic' data types in the C programming language.
Casting and dereferencing void pointers increases the risk of runtime errors that can be detected at compile time in other languages.

Phoenix++ addresses this by using template classes in its C++ framework implementation.
It takes a different approach by introducing modularity and the idea of containers and combiners, having the effect of embedding the user code at the heart of the framework.
However, there is an assumption that the user is aware of the available containers and the best selection is known before compilation.
An intimate understanding of the internal workings of the framework is required if a new container is needed for an application.
Moreover, some configurations require tuning at compile time restricting the data size at runtime.
In both these frameworks the development of optimizations impacts the programmability of the framework.
The objective in implementing a framework in Java is to eliminate the need for the user to write code beyond the functionality of the application; addressing the programmability and assessing the performance.

\subsection{MapReduce for Java (MR4J)}

To evaluate the capabilities of MapReduce on the JVM, MR4J has been developed.
The design principles behind MR4J are:

1) To maximize the use of standard Java libraries and exclude the use of native code to maintain portability across hardware architectures and operating systems.

2) To create a minimal API and return to the simplicity of the original Google implementation of MapReduce in order to encourage the user to concentrate on algorithmic development rather than ad-hoc parallelization.

3) To keep the implementation simple and encapsulate the internal working of the framework exposing only the fundamental API elements.

4) To target productivity while assessing performance in a transparent (to the programmer) manner with the implemented integrated optimizer.

At the center of MR4J's design are two elements, the scheduler and the collector of intermediate (key, value) pairs.
The \texttt{ForkJoinPool} class introduced in JDK 1.7 provide a clean, off-the-shelf scheduler focusing on lightweight tasks executing on worker threads accessed from a work-stealing queue \cite{Lea:2000}.
This compares to the scheduling approach of Phoenix and removes the need to implement a new scheduler.
In the existing frameworks the collection of intermediate (key, value) pairs is local to each worker thread and not directly transferable to Java tasks.
Phoenix demonstrates the flexibility of using a hash table for that purpose and MR4J selected the same approach.
Once the map phase is complete the values are passed as an argument into the reduce method as a \texttt{List} interface for user manipulation.

\section{Optimization}

\begin{figure*}
\centering
\includegraphics[trim = 0mm 7mm 0mm 5mm, clip, width=145mm]{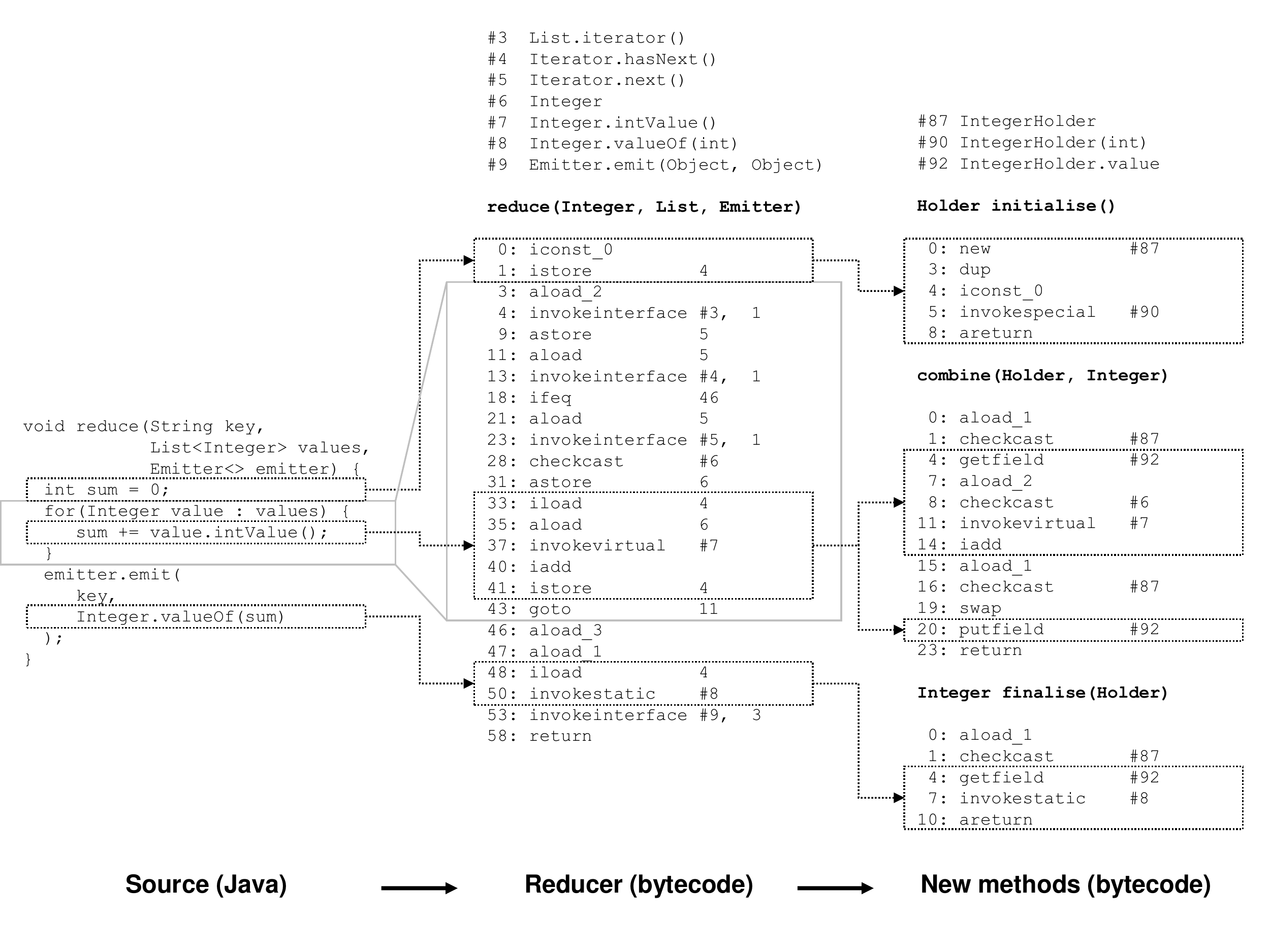}
\caption{Transformation of the reduce method for a word count application using MR4J.}
\label{fig:transformation}
\end{figure*}

The concept of a \textit{combiner} method to improve the locality, while reducing data, was first introduced in the original Google MapReduce framework \cite{Dean:2008}.
Its purpose is to combine emitted values locally on a processing node in order to limit the data transferred before and during the reduce phase.
In the multicore implementations, with direct access to all (key, value) pairs, it is possible to eliminate the reduce phase altogether.
Figure \ref{fig:mapcombine} illustrates how the word counting example can achieve this with a simple accumulator (an initial value of zero is assumed).

In related frameworks this optimization is manual and it is under the responsibility of the user to implement it.
Various combine and reduction algorithms have similar characteristics to the one explored in this example and therefore they can benefit from the automatic optimization explored in this paper.
This improvement will:

1) limit the source code written;

2) reduce the possibility of errors; and

3) improve performance of benchmarks where a combine method is feasible but not implemented.

The dynamic compiler is not able to optimize this case due to the semantic distance between the map and reduce methods.
They run in two phases of operation and are both embedded in tasks running in distinct time frames.
Consequently, the dynamic compiler will never see the interaction between the generation and reduction of intermediate values.
The developed optimizer is aware of this fact and by re-writing bytecode enacts the dynamic compiler to further improve the generated machine code, completely transparently to the user.

\subsection{MR4J Modifications}

Figures \ref{fig:mapreduce} and \ref{fig:mapcombine} illustrate the desired transformations in the context of MR4J in order to replace the reduce execution flow with combining.
The primary change is to provide an intermediate (key, value) pair collector that is aware of combining values (the intermediate value is held in a private encapsulating object (a \texttt{Holder})).
The same collector strategy is employed, the thread-safe hash table, with a different implementation of the emitter interface provided to the map method. Originally a new key would instantiate a new list to collect values.
In the optimized execution flow, a new key will instantiate a new holder and the value will be combined with the intermediate value held.
Before the results are returned to the user a finalization method will convert the intermediate value into the resulting value.

\subsubsection{Runtime Transformation}

The transformation of code during class loading is detailed in Figure \ref{fig:transformation}.
The reduce method is analyzed to create an intermediate representation that identifies three code fragments that will map onto the three methods required to implement the combiner in MR4J.
The purpose of each generated method is:

\texttt{Holder initialize();} provides an initial intermediate representation for values as a holder type.
In the case of all types it will provide a mutable boxing class.

\texttt{void combine(Holder, V);} contains the code from the reduce method that implements the combining.
The mutable value in the holder is modified to include the information required from the emitted value.

\texttt{V finalize(Holder);} converts the intermediate representation of the value into its final form.

Due to the implementation of Java generics, the \texttt{combine} and \texttt{finalize} methods also have a generated synthetic bridge method to act as an interface due to type erasure.
The methods ensure that type information is not erased from user code and the correct type is associated with objects on the stack during execution.
These have been omitted from Figure \ref{fig:transformation} for brevity.

The transformation is applicable when two conditions are satisfied.
Firstly, the reducer iterates over all intermediate values.
Secondly, the reduce operation is dependent only on the current intermediate value and current value in the iteration.
There are two idiomatic reducers handled directly in code that either use the size or first element in the intermediate value list.
Other complexities in determining correctness are provided by the MapReduce semantics and, therefore, they not need to be considered in the transformation.
For example should a value contain shared mutable state in a method executed, this must be thread-safe for the reduce method to provide a correct answer.
The implemented technique makes possible the potential analysis and implementation of verification code that provide hints at where violations to the safety of a MapReduce application lie.
The semantics of the framework add defined constraints that simplify checks that general purpose programming requires.

\subsection{Implementation}

A Java agent \cite{Binder:2007} was chosen as the most suitable technique to generate the new methods since it is simple to identify implementations of the reduce method.
The first step was to create an alternative execution flow in the MapReduce framework that uses the generated
methods that are hidden from the user, i.e. they contain no functionality and cannot be accessed or overriden outside of the declared package.
When the class loader loads the reduce class, it rewrites the access to these methods so they can be overridden at runtime.
The process of transforming the code follows the steps below:

1) Parse the reduce method to create an intermediate representation of the code in a program dependency graph.

2) Identify the conditions of the loop iterating over the values ensuring coverage of all values.

3) Test that the initialization block contains no external data dependencies, determine the holder type required ,and copy adjusted bytecodes to the initialize method body.

4) Test the value iteration loop body for data dependencies (assuming that the operation is associative due to the semantics of the MapReduce framework).
Copy adjusted bytecode to the combine method body.

5) Identify the original bytecode relating to the finalization of the intermediate value, from the preparation of the stack for the emit method call.
Copy adjusted bytecode to the finalize method body.

6) Set the flag to return a constant of true rather than false to enable the optimized combining execution flow in the MR4J implementation.

\section{Performance Evaluation}

MR4J is evaluated in two stages.
The first stage explores:
a) the scalability of MR4J on two different hardware configurations, and
b) the comparative evaluation of MR4J against mature and hand optimized state-of-the-art implementations in C and C++; Phoenix and Phoenix++ respectively.
The second stage evaluates the performance benefits generated by the MR4J aware optimizer.

\subsection{Experimental Set-up}

\subsubsection{Hardware Platforms}

The experiments run on two different hardware platforms in order to explore the performance on a multicore \textit{workstation} and a larger NUMA multi-socket, multicore \textit{server}.
Table \ref{tab:configuration} presents the hardware and software configurations used during the evaluation.

\subsubsection{MapReduce Software Frameworks}

The evaluation compares MR4J against the hand-tuned Phoenix \cite{Yoo:2009} and Phoenix++ \cite{Talbot:2011} implementations.
These are both configured manually using hardware specific parameters; e.g. the size of L1 cache and the number of desired threads.
MR4J uses the same L1 cache size as its buffer size and the JVM is configured to use the default garbage collector (Parallel) with an initial and maximum heap size of 12GB.
Furthermore, the \texttt{-XX:+UseNUMA} flag is set for the server configuration.
Each benchmark is executed ten times (Java includes a five iteration warm-up) and the average execution time is used to report results.

\subsubsection{Benchmarks}

The benchmarks distributed and used by Phoenix and Phoenix++ have been ported and validated on MR4J for a fair comparison.
The benchmark suite consists of the following applications, as detailed by Yoo \textit{et~al.} \cite{Yoo:2009}: Histogram (HG), K-Means Clustering (KM), Linear Regression (LR), Matrix Multiply (MM), Principal Component Analysis (PC), String Match (SM), and Word Count (WC).

\noindent In order to ensure that the same algorithms are executed across all three frameworks, modifications have been made to the original benchmarks.
For Histogram, Phoenix++ iterates over individual pixels; however due to performance and memory constraints, Phoenix and MR4J iterate over chunks of data, emitting values after partial combination in the map method.
Histogram and Word Count omit the requirement to sort the keys as this is testing the efficiency of parallel sorting algorithms rather than the core of MapReduce.

\begin{table}
\centering
{\small
\begin{tabular}{ p{26mm} C{23mm} C{23mm} }
\hline
                 & \textbf{Workstation} & \textbf{Server}       \\
\hline
Processor        & Intel Core i7        & AMD Opteron           \\
                 & 4770 3.4GHz          & 6276 2.3Ghz           \\
Cores            & 4                    & 64 (4 x 16)           \\
Hardware threads & 8                    & 64                    \\
L1 Cache         & 32kB per core        & 16kB per core         \\
L2 Cache         & 256kB per core       & 2MB per 2 cores       \\
L3 Cache         & 8MB per 4 cores      & 8MB per 8 cores       \\
Main memory      & 16GB                 & 252GB                 \\         
\hline
OS             & Windows 8.1            & Ubuntu 12.04          \\
C/C++ compiler & gcc 4.8.3              & gcc 4.6.4             \\
Java           & \multicolumn{2}{c}{Java SE 1.8.0\_20}          \\
JVM            & \multicolumn{2}{c}{Java HotSpot 64-Bit Server} \\
               & \multicolumn{2}{c}{(build 25.20-b23)}          \\
\hline
\end{tabular} }
\caption{Hardware and software configurations.}
\label{tab:configuration}
\end{table}

\begin{table}
\centering
{\small
\begin{tabular}{ C{6mm} p{43mm} C{11mm} C{11mm} }
\hline
   & \textbf{Dataset}                    & \textbf{Keys}  & \textbf{Values} \\
\hline
HG & 1.4GB 24-bit bitmap image           & Medium         & Large           \\
KM & 500,000 3-d points (100 clusters)   & Small          & Large           \\
LR & 3.5GB file                          & Small          & Large           \\
MM & 3,000 x 3,000 integer matrices      & Medium         & Medium          \\
PC & 3,000 x 3,000 integer matrix        & Medium         & Medium          \\
SM & 500MB key file                      & Small          & Small           \\
WC & 500MB text document                 & Large          & Large           \\
\hline
\end{tabular} }
\caption{Benchmark Input Data.}
\label{tab:dataset}
\end{table}

The benchmarks demonstrate a variety of workloads, inputs, intermediate and output results.
All of these benchmarks, originally from the Phoenix paper \cite{Yoo:2009}, contain combiner methods.
These combiners are all generated by the optimizer described in this paper.
The challenge for all three frameworks was to generate a combiner for the K-Means Clustering benchmark as it requires state to obtain the average (e.g. the total number of points in a cluster).
In this case the combiner or the intermediate value contain the running sum of point coordinates.
The sum is normalized in the reducer for MR4J or in the main body of the application for Phoenix and Phoenix++.
Table \ref{tab:dataset} presents the input data sets with an approximate categorization of key and value counts.

\subsection{Performance Results}

\begin{figure}
\centering
\includegraphics[trim = 19mm 23mm 22mm 20mm, clip, width=80mm]{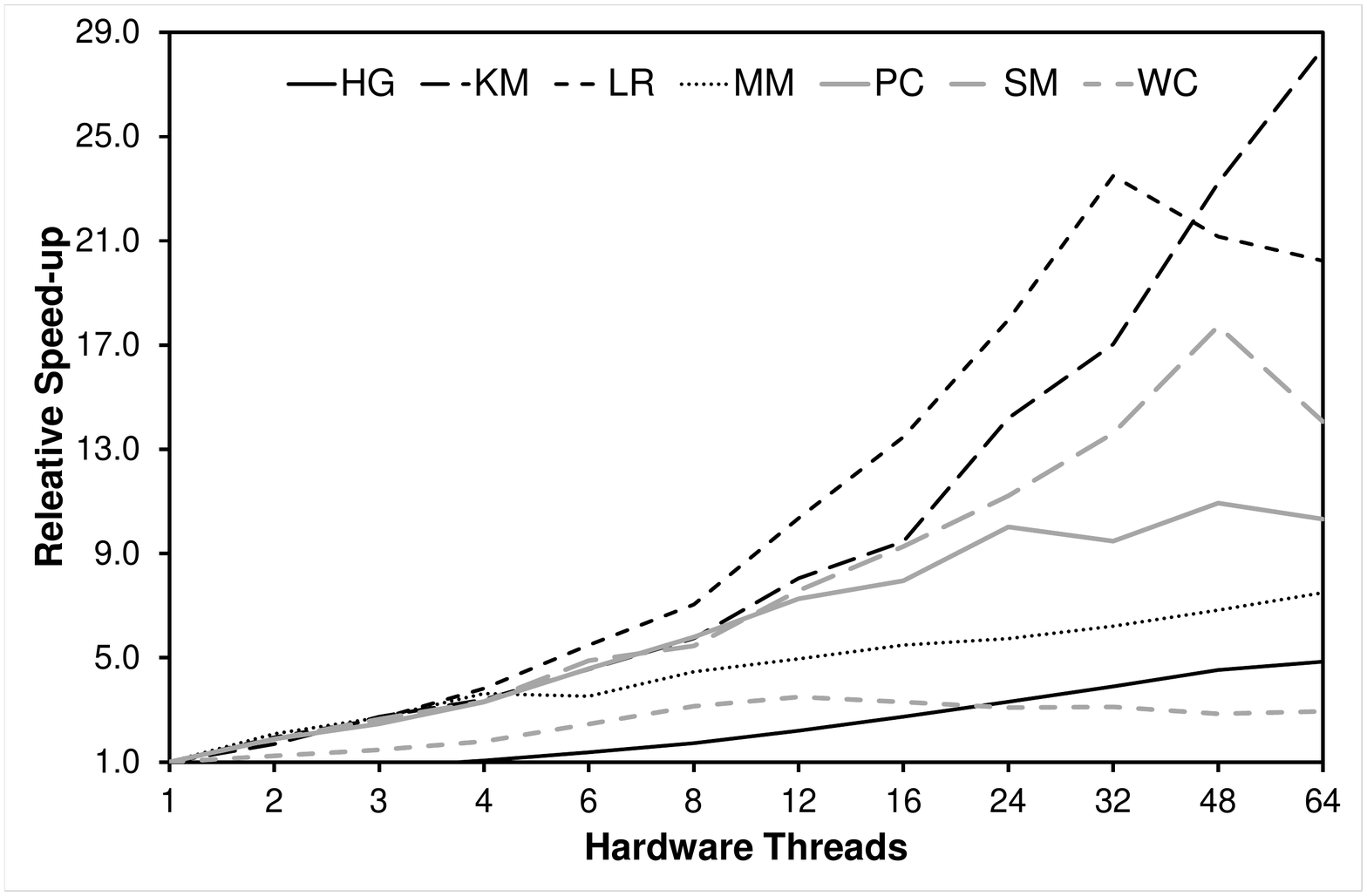}
\caption{MR4J scalability on the server configuration (one thread as the baseline).}
\label{fig:benchmarks}
\end{figure}

The scalability of MR4J can be seen in Figure \ref{fig:benchmarks} for the server configuration.
Having as a baseline the execution time on one core, the workstation shows a consistent scalability over all hardware threads, with an average of 2.85 on four cores and 3.73 on all eight hyperthreads.
Regarding the scalability of MR4J on the server configuration (Figure \ref{fig:benchmarks}), three groups of performance can be observed depending on their compute intensity and overhead of (key, value) pair generation summarized in Table \ref{tab:dataset}.

\begin{figure}
\centering
\includegraphics[trim = 19mm 23mm 22mm 20mm, clip, width=80mm]{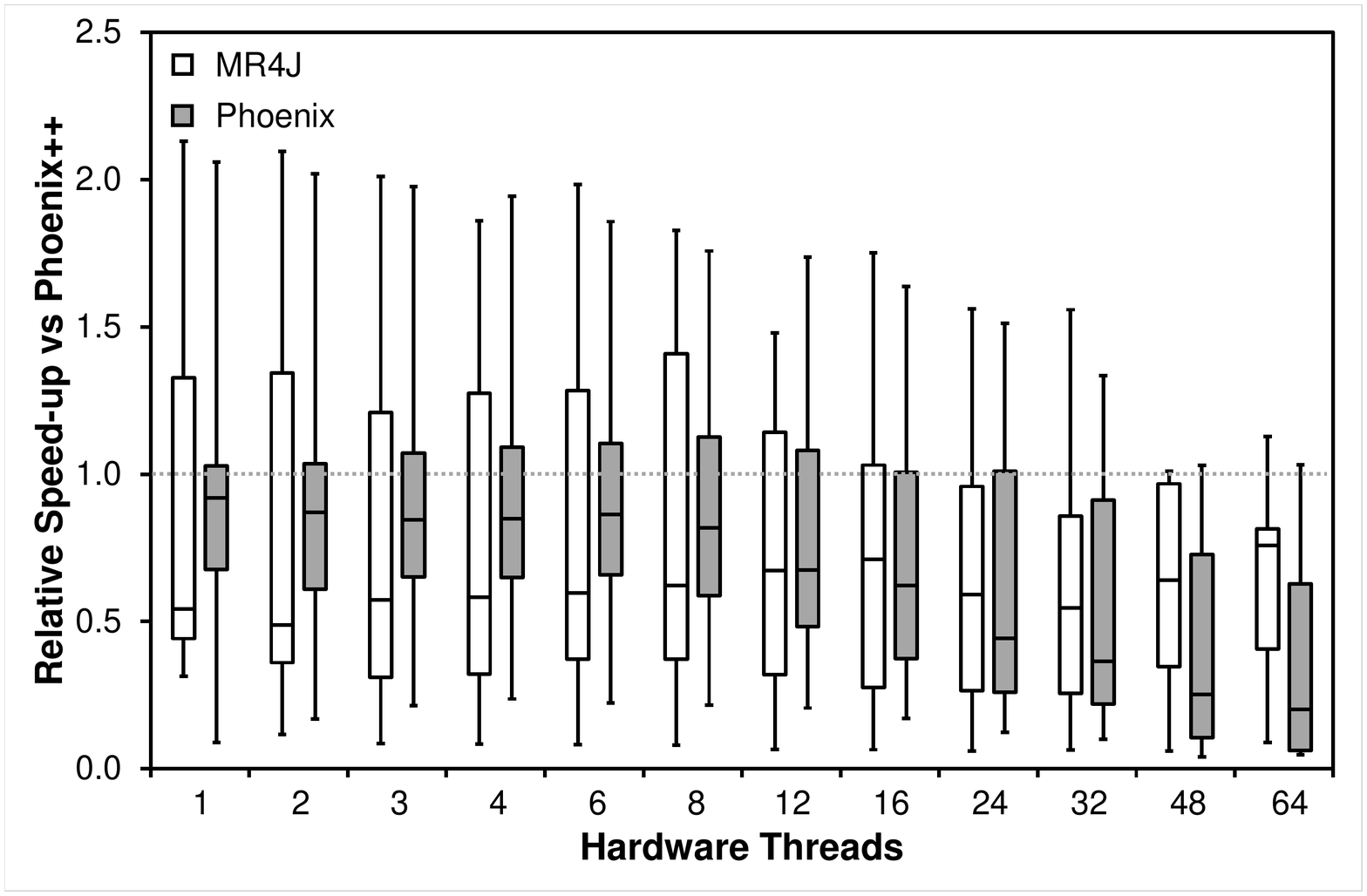}
\caption{Relative speedup of Phoenix and MR4J against Phoenix++ on server(higher is better).}
\label{fig:speedup}
\vspace{-3mm}
\end{figure}

\begin{figure}
\centering
\includegraphics[trim = 19mm 23mm 22mm 20mm, clip, width=80mm]{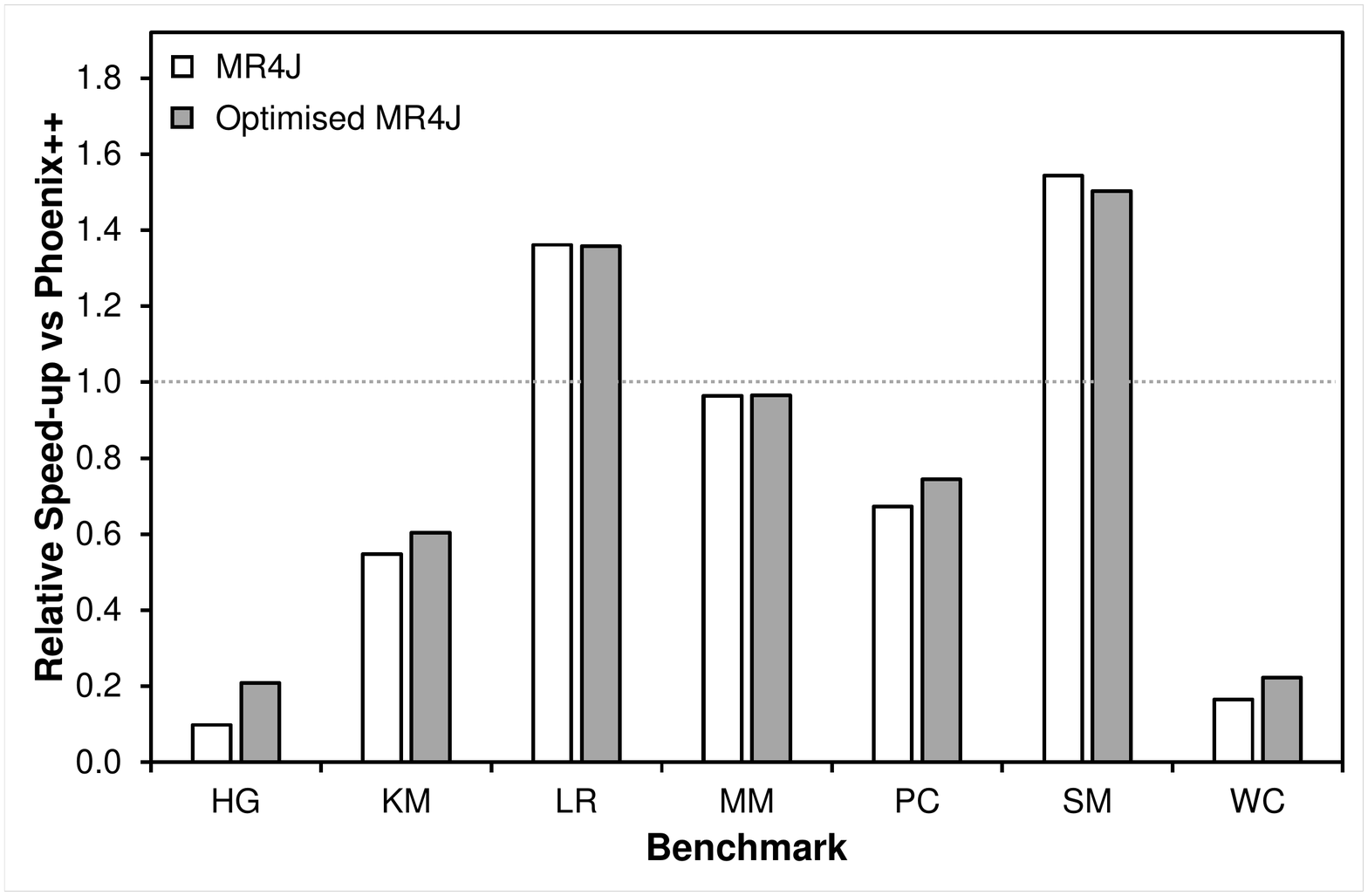}
\caption{MR4J per-benchmark speedup with and without the optimizer relative to Phoenix++ on server.}
\label{fig:optimization}
\vspace{-3mm}
\end{figure}

Figure \ref{fig:speedup} contains the speedup of MR4J and Phoenix relative to Phoenix++ on the server configurations respectively.
Furthermore, Figure \ref{fig:optimization} take a more fine-grain approach and illustrate the relative speedup of MR4J against the top-performing Phoenix++ with and without the implemented optimizer respectively per benchmark.
Regarding the workstation configuration, a consistent performance behavior can be observed between MR4J, Phoenix and Phoenix++.
The performance falls in-between the two hand-tuned frameworks with the median around 0.66 for MR4J and 0.39 for Phoenix for all hardware thread counts.
The server configuration reveals a different set of results illustrating the challenges of developing scalable software for multisocket NUMA architectures.
When using the same socket (1--16 threads) the performance of MR4J and Phoenix is comparative to Phoenix++ which consistently out-performs them (0.61 and 0.81 respectively).
Scalability was a primary objective in the development of Phoenix++ \cite{Talbot:2011} and the results are supported by this evaluation.
The NUMA aware setting in the JVM is able to maintain a consistent level of performance, unlike Phoenix which employs only its locality optimizations.
However, the speedups of MR4J and Phoenix are 0.76 and 0.20 compared to Phoenix++ when using all hardware threads.

\subsection{Optimization Performance}

Figure \ref{fig:optimization} illustrates the relative speedup of MR4J against Phoenix++ before and after the optimizer is enabled for each of the benchmarks.
The majority of the benchmarks on both configurations show a significant speedup, and thus, closing the gap between MR4J and Phoenix++.
String Match is an exception, exposing the overheads of instantiating and maintaining the intermediate value.
This is due to the nature of the benchmark which has few keys, few values and little computation that can be optimized.
The main overheads of the optimizer are when detecting classes that extend the Reducer and then generating the combining code.
Since the optimizer instruments every Java class, the effect on the detection and transformation times are, on average per class, 81$\mu$s and 7.6ms respectively, which is negligible in comparison to the execution time of the benchmarks.

\section{Discussion}

The introduced MR4J is a lightweight MapReduce framework based on the standard JDK classes.
By using a simple API and by utilizing Java interfaces it is possible to improve the framework while maintaining the backwards compatibility ethos of Java.
The presented optimization illustrates how a single map method can be used in two alternative execution flows, one to reduce values and the other to combine them, thanks to the use of the \texttt{Emitter} interface.

On a multicore architecture, MR4J provides consistently better execution times than the hand-optimized C equivalent and, after optimization, is within reach of the equivalent in C++.
Phoenix and Phoenix++ offer powerful and scalable tools but with more complicated APIs that require manual configuration and tuning.
The benchmarks where MR4J is superior are those where data is organized in arrays.
The dynamic compiler is able to better optimize array accesses (through the automatic memory manager) than pointer arithmetic alone in a static compiler.
However, in benchmarks where heavy object creation is required, the ability of C and C++ to cast directly to data highlights the overhead of object allocation and management in Java.
K-Means Clustering with Points and Word Count with Strings are such examples.

\begin{figure}
\centering
\includegraphics[trim = 19mm 23mm 22mm 20mm, clip, width=80mm]{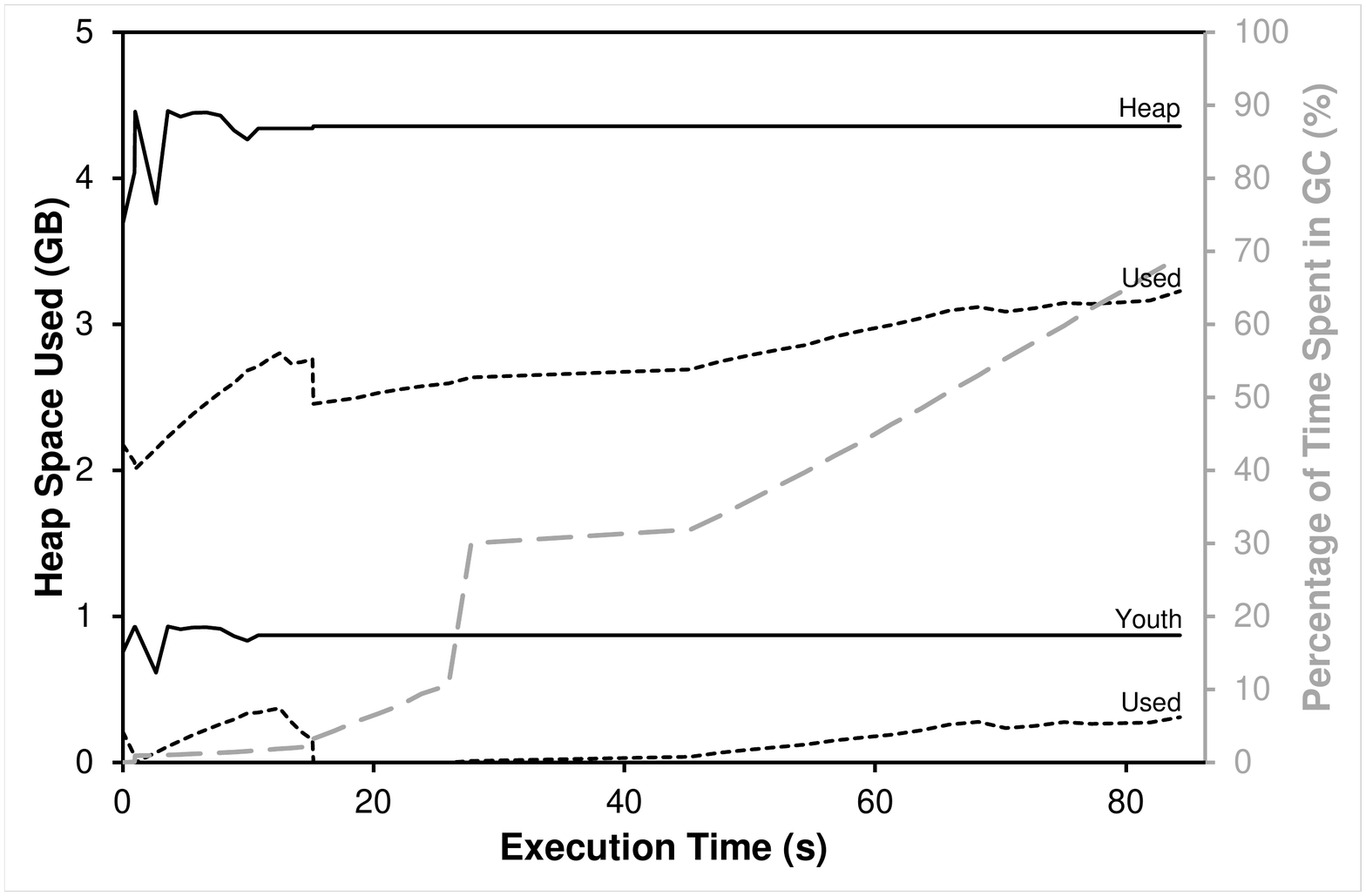}
\caption{Word Count on MR4J: Heap usage and percentage of runtime spent in garbage collection.}
\label{fig:gc}
\vspace{-3mm}
\end{figure}

The optimization presented in this paper changes the execution flow within the framework. Borrowing the notion of manual combining from existing MapReduce frameworks, the implemented optimizer automates this process at runtime.
The optimizer uses the semantics of the framework and the structure of user code to eliminate the reduce phase and combine intermediate values as they are emitted from the map method.
This has the effect of improving the execution time for the majority of the tested benchmarks.

\begin{figure}
\centering
\includegraphics[trim = 19mm 23mm 22mm 20mm, clip, width=80mm]{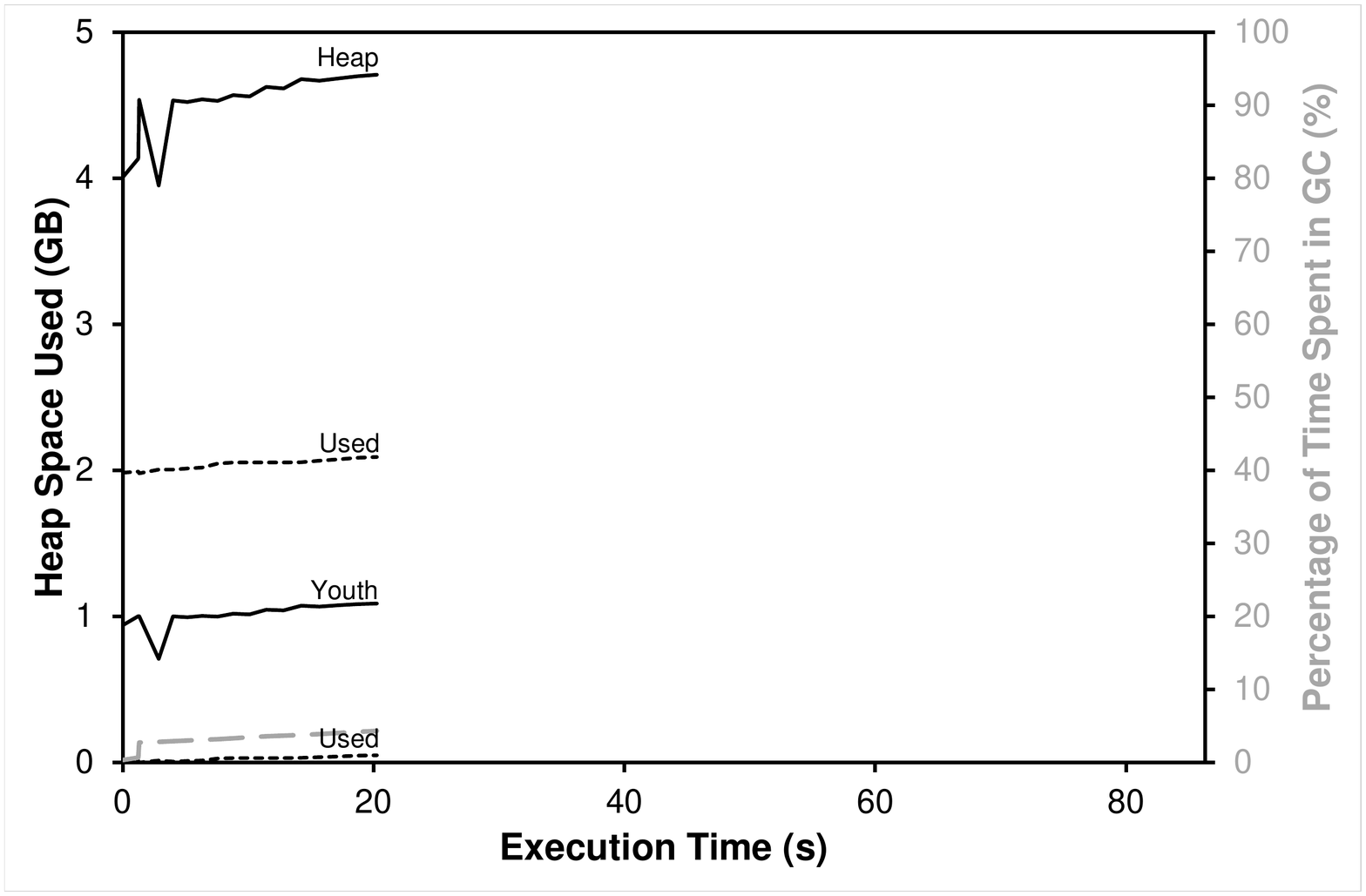}
\caption{Word Count on optimized MR4J: Heap usage and percentage of runtime spent in garbage collection.}
\label{fig:gcoptimized}
\end{figure}

The cause of the observed speedup is the improved interaction between the optimized executed code, the dynamic compiler and the Garbage Collector (GC).
Figures \ref{fig:gc} and \ref{fig:gcoptimized} visualize the heap usage for the word count application without and with the optimizer respectively.
The execution time axes are the same for a direct comparison.
The heap usage is similar for both configurations showing a noticeable and steady increase in the size of the heap used since more references are stored for the intermediate values.
The stark difference is in the secondary axis, the time spent in the GC.
Without the optimization the inefficiency lies in the fact that Java must maintain (i.e. keep in the heap) all the object references for the intermediate values generated during the map phase.
This results in their premature promotion into the older generations before they die (and collected during minor collections).
This, consequently, results in major collections that severely impact performance.
The optimization, in turn, increases performance by:

1) reducing the number of objects allocated which avoids unnecessary object promotions leading to major GC cycles;

2) improving execution time by omitting completely the reduce phase;

3) enabling the dynamic compiler to introduce additional scalar replacements, and

4) reducing the utilized heap size and, thus, enabling larger data sets to be used; increasing the potential for utilizing smaller Big Data jobs (as mentioned in the Hadoop job analysis \cite{Appuswamy:2013}).

\begin{figure}
\centering
\includegraphics[trim = 19mm 23mm 22mm 20mm, clip, width=80mm]{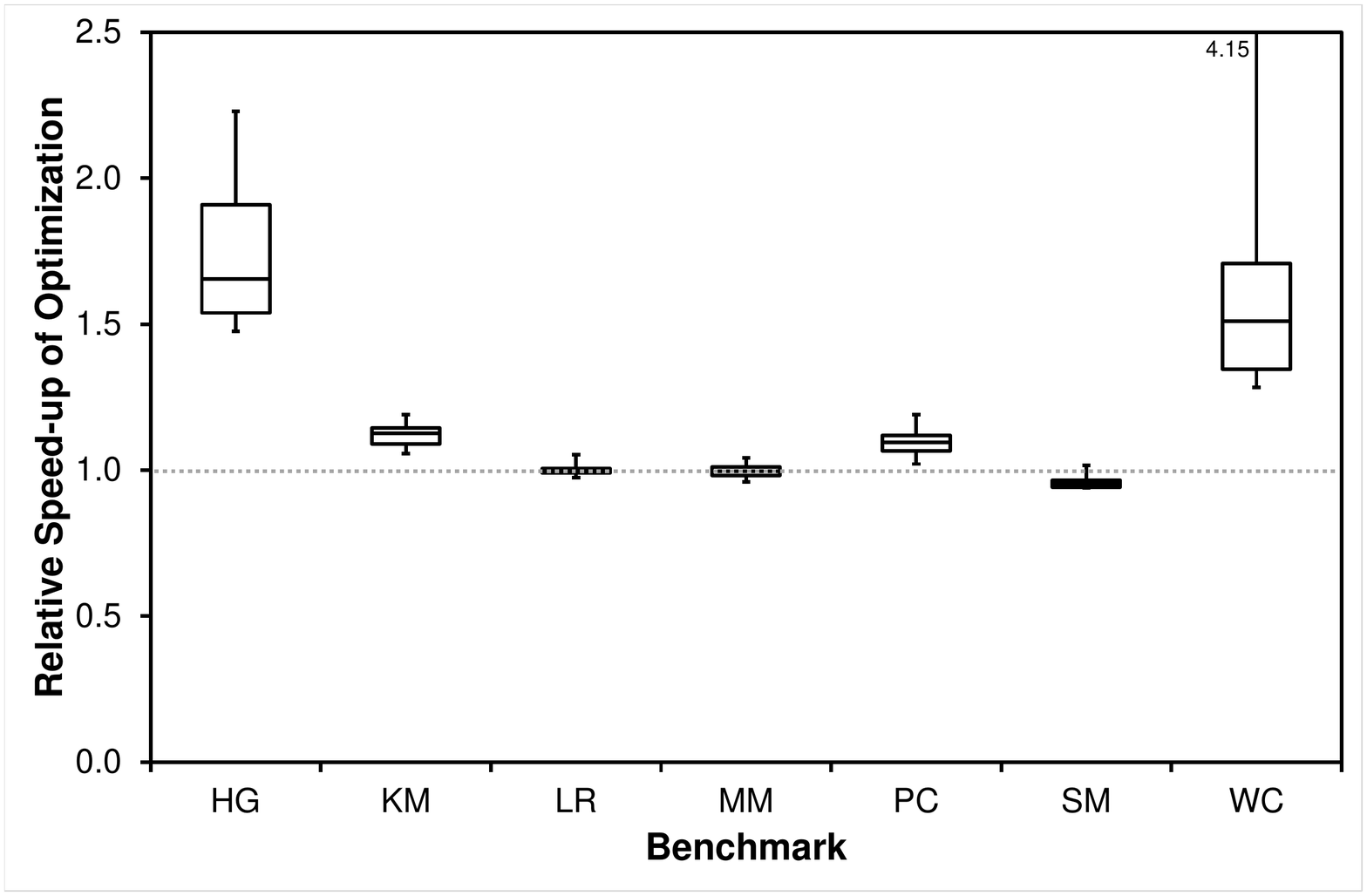}
\caption{Word Count on optimized MR4J: Heap usage and percentage of runtime spent in garbage collection.}
\label{fig:gcperformance}
\vspace{-3mm}
\end{figure}

The JVM, as publicly distributed by Oracle, contains a variety of GC algorithms allowing different tuning parameters and configurations.
Figure \ref{fig:gcperformance} depicts the relative (to the baseline un-optimized version) speedup of each benchmark when all the combinations of GC algorithms, heap sizes, and number of hyper threads are averaged.
The figure also shows that the benchmarks with the greatest reliance on (key, value) pairs (HG and WC) are improved the most.
String Match has four keys with 910 values; whereas Histogram has 768 keys and $1.4\times 10^9$ values.

\section{Conclusions}

This paper introduces MR4J, a lightweight Java based MapReduce framework for shared-memory multicore architectures built on standard JDK classes.
MR4J focuses on ease-of-programmability via a simple API in contrast to equivalent frameworks where performance is extracted via complicated manual tuning required by the programmer.
The performance loss, due to its simplicity, is overcome by a novel optimizer built for the framework.
The optimizer exploits semantic information inherently contained within the parallel software framework transparently to the user.
The design of MR4J aims to either supplement developers of large MapReduce algorithms, improve productivity or simply execute smaller applications.

The performance of MR4J is comparative to the equivalent state-of-the-art Phoenix framework, written and hand-optimized in C.
Thanks to the expressiveness, type safety and portability of Java, it creates a more productive and portable framework with comparative performance.
The original implementation of MR4J was positioned in between the two state-of-the-art MapReduce frameworks, Phoenix and Phoenix++, performance wise.
The lack of a combiner phase was penalizing performance and therefore the optimizer was implemented to supplement the framework.
The presented co-designed optimizer automates the, previously hand-optimized, combining phase in order to improve performance.
Without any modifications to user code, the optimized MR4J improves its performance up to 2.0x bridging the gap from the manually-tuned Phoenix++ to just 17\%.

The work presented in this paper is a proof-of-concept that if semantic information can be passed from the application developer to the parallel framework and the compiler, significant performance improvements can be achieved.
Especially nowadays, with the advent of complex multi-layered Big Data frameworks that are deployed on top of diverse and often heterogeneous hardware resources, semantic-based optimizations will be even harder to achieve.
In the quest for achieving vertical co-designed optimizations we plan to exercise this and other developed optimizations directly into the underlying compiler.
To that end, we plan to augment the existing state-of-the-art Graal compiler \cite{Duboscq:2013} with semantically enriched hooks in order to transfer the necessary information from the application to the compiler.
The formalization of the information flow from the application level down to the compiler and runtime level is of paramount importance in order to bridge the semantic gaps both between different software frameworks and between software and hardware.

\section{Acknowledgments}

This research was conducted with support from the UK Engineering and Physical Sciences Research Council (EPSRC), on grants AnyScale Apps EP/L000725/1, DOME EP/ J016330/1 and PAMELA EP/K008730/1.
Mikel Luj{\'a}n is supported by a Royal Society University Research Fellowship.

\bibliographystyle{abbrv}
\bibliography{paper}

\end{document}